\documentclass[conference]{IEEEtran}
\IEEEoverridecommandlockouts
\usepackage{cite}
\usepackage{amsmath,amssymb,amsfonts}
\usepackage{algorithmic}
\usepackage{graphicx}
\usepackage{subfig}
\usepackage{textcomp}
\makeatletter

\newcommand{\Rmnum}[1]{\expandafter\@slowromancap\romannumeral #1@}
\makeatother
\usepackage{multirow, makecell}
\def\BigRoman{\uppercase\expandafter{\romannumeral\number\count 255 }}
\def\Romannumeral{\afterassignment\BigRoman\count255=}
\usepackage{tikz}
\def\BibTeX{{\rm B\kern-.05em{\sc i\kern-.025em b}\kern-.08em
    T\kern-.1667em\lower.7ex\hbox{E}\kern-.125emX}}

\begin{document}

\title{Classification of Tactile Perception and Attention on Natural Textures from EEG Signals
\footnote{{\thanks{20xx IEEE. Personal use of this material is permitted. Permission
from IEEE must be obtained for all other uses, in any current or future media, including reprinting/republishing this material for advertising or promotional purposes, creating new collective works, for resale or redistribution to servers or lists, or reuse of any copyrighted component of this work in other works.

Research was funded by Institute of Information \& Communications Technology Planning \& Evaluation (IITP) grant funded by the Korea government (No. 2017-0-00451, Development of BCI based Brain and Cognitive Computing Technology for Recognizing User’s Intentions using Deep Learning; No. 2019-0-00079, Artificial Intelligence Graduate School Program (Korea University)).}
}}
}

\author{\IEEEauthorblockN{Myoung-Ki Kim}
\IEEEauthorblockA{\textit{Dept. Artificial Intelligence} \\
\textit{Korea University}\\
Seoul, Republic of Korea \\
\textit{LG Display}\\
kim\_mk@korea.ac.kr}
\and
\IEEEauthorblockN{Jeong-Hyun Cho}
\IEEEauthorblockA{\textit{Dept. Brain and Cognitive Engineering} \\
\textit{Korea University}\\
Seoul, Republic of Korea \\
jh\_cho@korea.ac.kr}
\and
\IEEEauthorblockN{Ji-Hoon Jeong}
\IEEEauthorblockA{\textit{Dept. Brain and Cognitive Engineering} \\
\textit{Korea University}\\
Seoul, Republic of Korea \\
jh\_jeong@korea.ac.kr}}




\maketitle

\begin{abstract}
Brain-computer interface allows people who have lost their motor skills to control robot limbs based on electroencephalography. Most BCIs are guided only by visual feedback and do not have somatosensory feedback, which is an important component of normal motor behavior. The sense of touch is a very crucial sensory modality, especially in object recognition and manipulation. When manipulating an object, the brain uses empirical information about the tactile properties of the object. In addition, the primary somatosensory cortex is not only involved in processing the sense of touch in our body but also responds to visible contact with other people or inanimate objects.
Based on these findings, we conducted a preliminary experiment to confirm the possibility of a novel paradigm called touch imagery.
A haptic imagery experiment was conducted on four objects, and through neurophysiological analysis, a comparison analysis was performed with the brain waves of the actual tactile sense. Also, high classification performance was confirmed through the basic machine learning algorithm.
\end{abstract}

\begin{small}
\textbf{\textit{Keywords-brain-computer interface; electroencephalography; neurohaptics; touch imagery; haptic imagery; tactile perception; haptic perception}}\\
\end{small}

\section{Introduction}
Brain-computer interface (BCI) provides a means of communication for healthy users and users who have lost their motor skills due to several causes such as spinal cord injury or stroke\cite{kwak2017convolutional, yeom2014efficient, lee2015subject, won2017motion}. In addition, it enables device control by reflecting the user's intention through communication with the user's brain signals and various computers and machines\cite{kim2018commanding, jeong2020multimodal, lee2020continuous}. Electroencephalography (EEG) signal is non-invasive and has a high temporal resolution, which makes them most commonly used in BCI systems\cite{lee2018high, lee2020sessionnet, chen2016ITIFS}.

Most BCIs are guided only by visual feedback and do not have somatosensory feedback, an important component of normal motor behavior\cite{Johansson2009NatureReviewsNeuroscience, Abraira2013Neuron}.
Human has several sensory channels. Among them, the sense of touch is a very important sensory modality, especially in object recognition and manipulation. Various types of mechanoreceptors are distributed in the hand, providing detailed tactile information about contact events to the brain through afferent nerves. It also provides sophisticated sensitivity to the shape and surface properties of objects. Mechanoreceptors that sense tactile include slowly adapting type I (SA I) and rapidly adapting (RA) and Pacinian mechanoreceptors (PC). Features such as the shape of the objects are reflected in the activated spatial pattern of the SA I and RA fibers, which are densely located at the fingertips. The dynamic perception of fine natural textures depends on the transformation of high-frequency vibrations by RA and PCs\cite{Saal2017PNAS, Johansson2004NatureNeuro}.

People with impaired tactile sensations struggle with daily activities due to a lack of information about the mechanical contact conditions that the brain needs to plan and control object manipulation. Visual modalities have been studied extensively from a perceptual and cognitive perspective. However, vision only provides indirect information about these mechanical interactions. Specifically, while manipulating objects, the brain uses sensory predictions and afferent signals to select and implement motion-step controllers that adjust motor power to the physical properties of the objects involved. Without tactile somatosensory feedback, simple tasks are clumsy and slow. The need for somatosensory feedback in BCI has long been suggested as the next step in completing upper limb restoration\cite{Nowak2003Brain, Ede2019NatureNeur}. In a recent study, Flesher et al and Ganzer et al. restored somatosensory feedback in invasive BCI and constructed a closed-loop system\cite{Ganzer2020Cell, Bouton2016Nature, Flesher2019bioXiv, Flesher2016SciMed}. As shown in these studies, the completion of the BCI control system is a closed-loop system including somatosensory feedback. However, few relevant non-invasive BCI studies have been reported. In the invasive BCI method, tactile information can be directly acquired and transmitted by implanting an electrode chip in the somatosensory region of the cerebral cortex. However, this method requires a surgical operation.

 In recent studies on tactile perception, we explored the possibility of using tactile feedback in non-invasive BCI. The primary somatosensory cortex is not only involved in processing the sense of touch in our body but also responds to visible contact with other people or inanimate objects\cite{Schirmer2019Cortex, Pisoni2018NeuroImage, Yao2017IEEE}. Focusing on this, we have come to think about how to acquire tactile information of an object with EEG without a direct tactile feedback device. In other words, we started research on how to use the empirical tactile information that human has. This study is the result of a preliminary experiment for this.

In this study, we designed a novel paradigm called touch imagery for decoding various brain signals corresponds to haptic perception. To the best of our knowledge,  this is the first study to show the possibility for the classification and analysis of natural haptic perception. We achieved high classification performance for 4-class touch imagery correspond to four different objects (fabric, glass, paper, and fur).

The rest of this document is organized as follows: Section {\Romannumeral 2} provides a description of the experimental protocol, EEG signal acquisition, and data analysis. Section {\Romannumeral 3} provides the results of the performance accuracy for class 4 classification and a discussion of our work. In session {\Romannumeral 4}, conclusion and future works are described.\\

\begin{figure}[t]
\centerline{\includegraphics[width = \columnwidth]{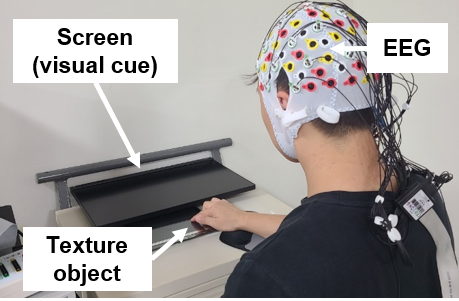}}
\caption{Experimental setup. Subjects were seated in a comfortable chair in front of the table. A screen was installed on the table so the subjects could see objects and visual cues. The subjects were tested with their right hand comfortably placed on the hand holder on the table.}
\end{figure}

\section {Materials and Methods}
\subsection{Participants}
Five healthy subjects with no history of neurological disease were recruited for the experiment (S1-S5; ages 27-39; five men; all right-handed). This study was reviewed and approved by the International Review Board, at Korea University [KUIRB-2020-0013-01], and written informed consent was obtained from all participants before the experiments.

\subsection{Experimental Setup}
During the experimental protocol session, subjects were seated in a comfortable chair in front of a 24-inch LCD monitor screen. A screen was installed on the table so the subjects could see objects and visual cues. The subjects were tested with their right hand comfortably placed on the hand holder on the table. Fig. 1 shows the experimental setup and environment during the entire session. The experiment consisted of two tasks. In the first ``Real Touch'' task, tactile recognition is performed by direct contact and sliding of the object's surface according to the signal, and in the second ``Touch Imagery'' task, we were asked to imagine the surface tactile characteristics of the objects according to the signal. During the experiment, subjects were asked to perform tactile perception and imagination of the four different objects shown in Fig. 2. Sufficient rest was given between each task. The experimental paradigm is shown in Fig. 3.

\subsection{Data Acquisition}
EEG data were collected at 2,500 Hz using 64 Ag/AgCl electrodes in 10/20 international system via BrainAmp (BrainProduct GmbH, Germany). At the same time, a 60 Hz notch filter was used to remove power frequency interference. The FCz and FPz were used as reference and ground electrodes, respectively. All impedances were maintained below 10 k$\Omega$.


\begin{figure}[t]
  \centering
  \subfloat(a)
  \begin{minipage}{.2\textwidth}\centering\includegraphics[width=0.9\columnwidth]{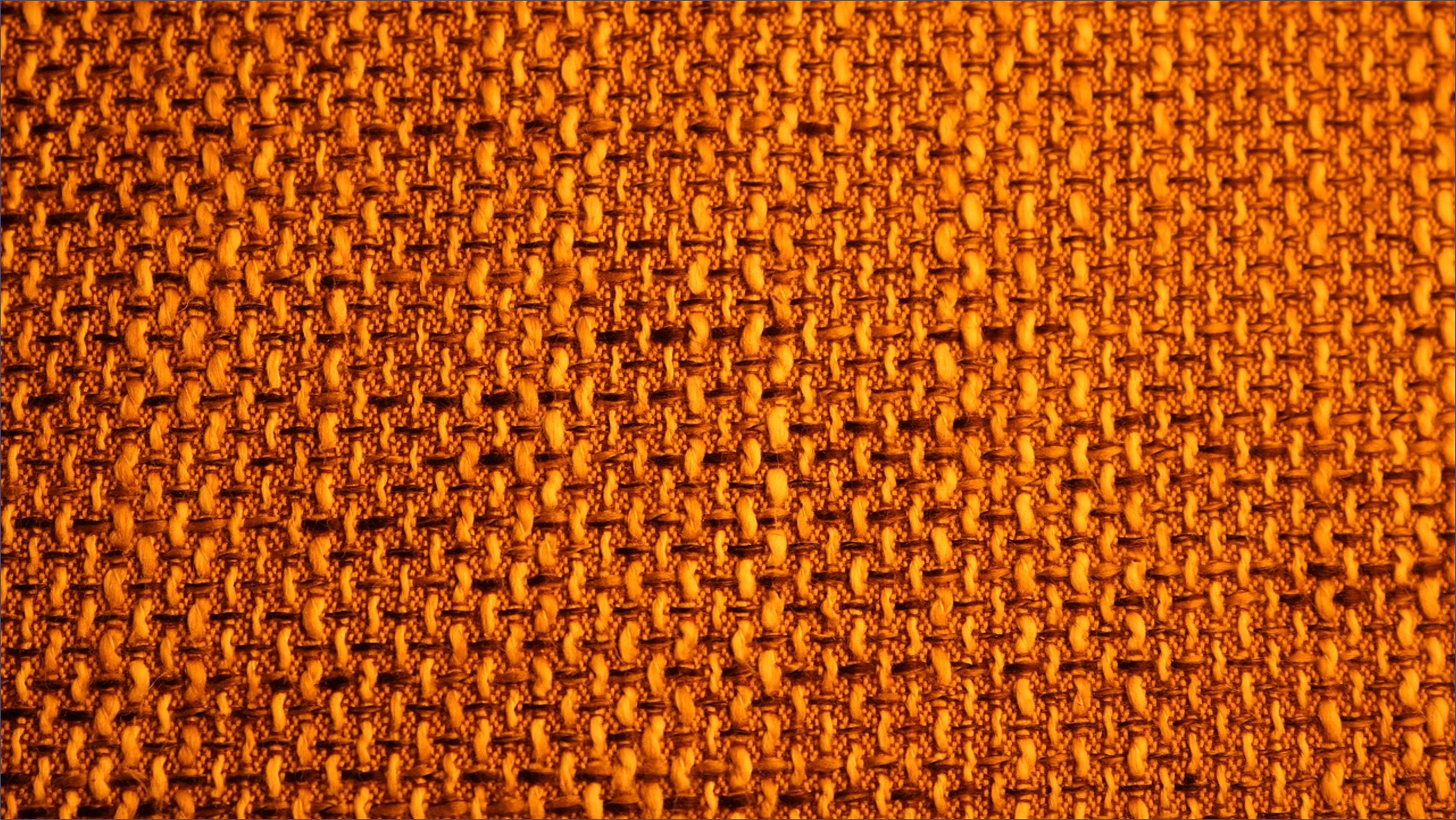}
  \end{minipage}
  \subfloat(b)
  \begin{minipage}{.2\textwidth}\centering\includegraphics[width=0.9\columnwidth]{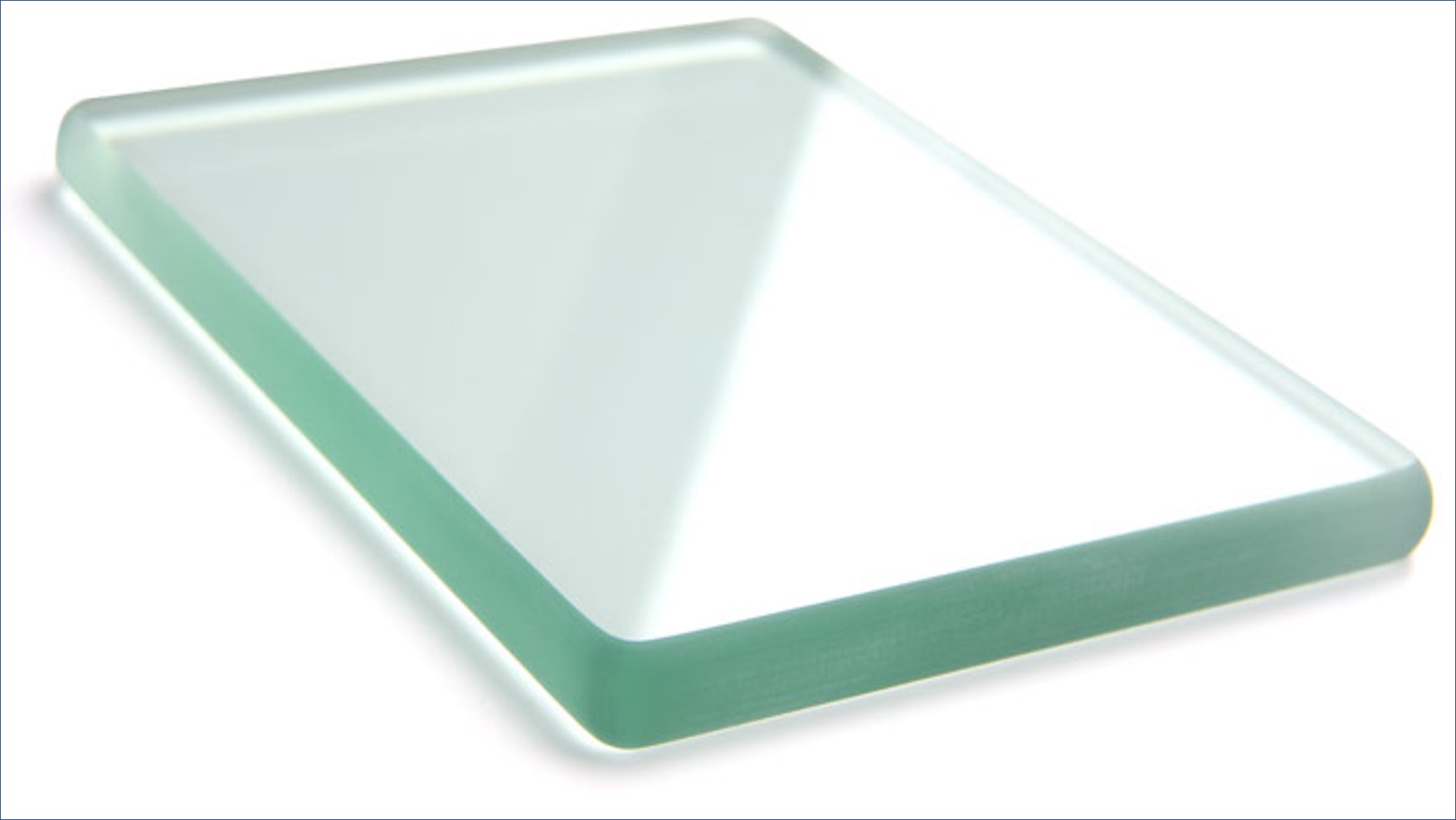}
  \end{minipage}
  \\
  \subfloat(c)
  \begin{minipage}{.2\textwidth}\centering\includegraphics[width=0.9\columnwidth]{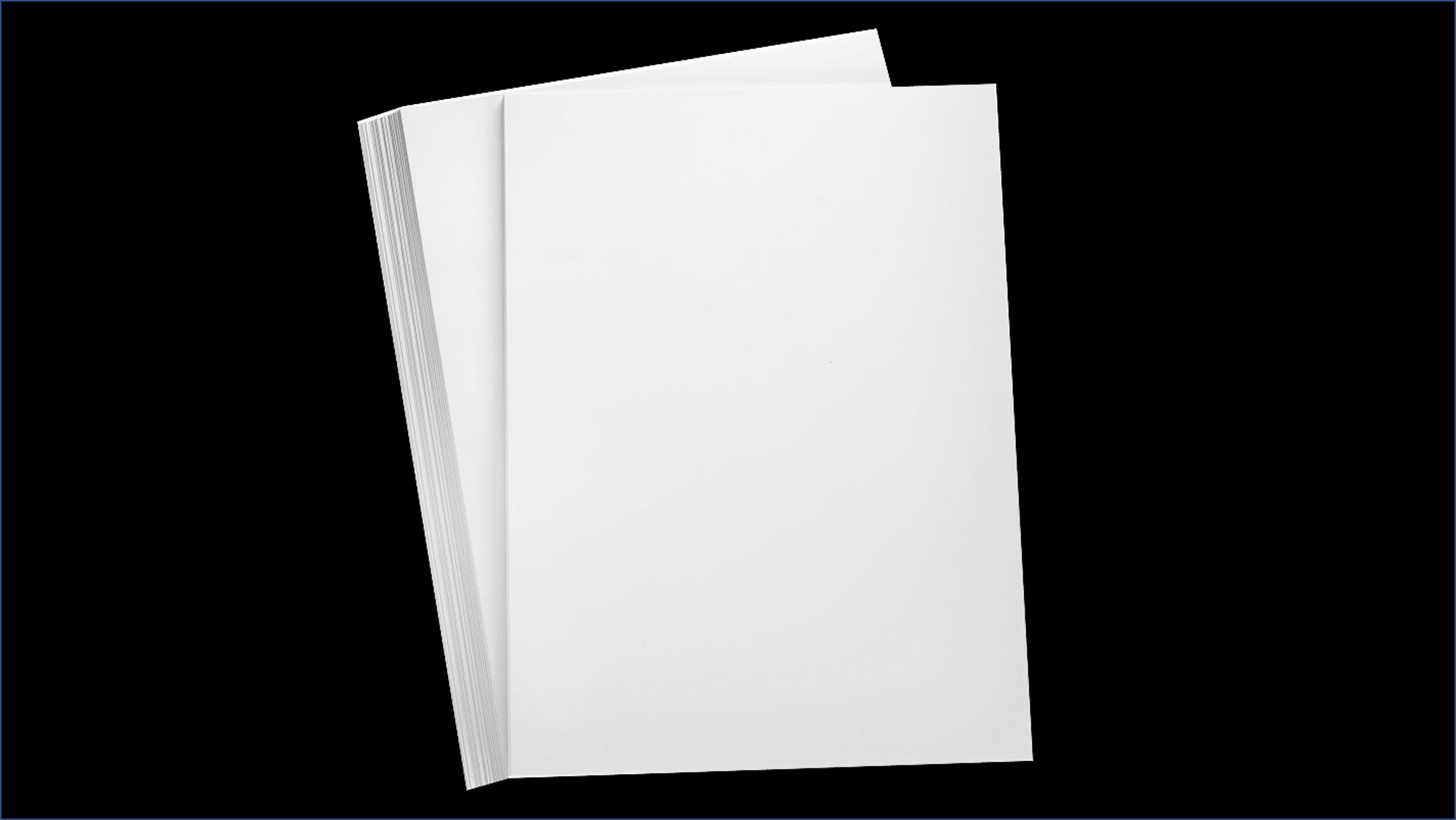}
  \end{minipage}
  \subfloat(d)
  \begin{minipage}{.2\textwidth}\centering\includegraphics[width=0.9\columnwidth]{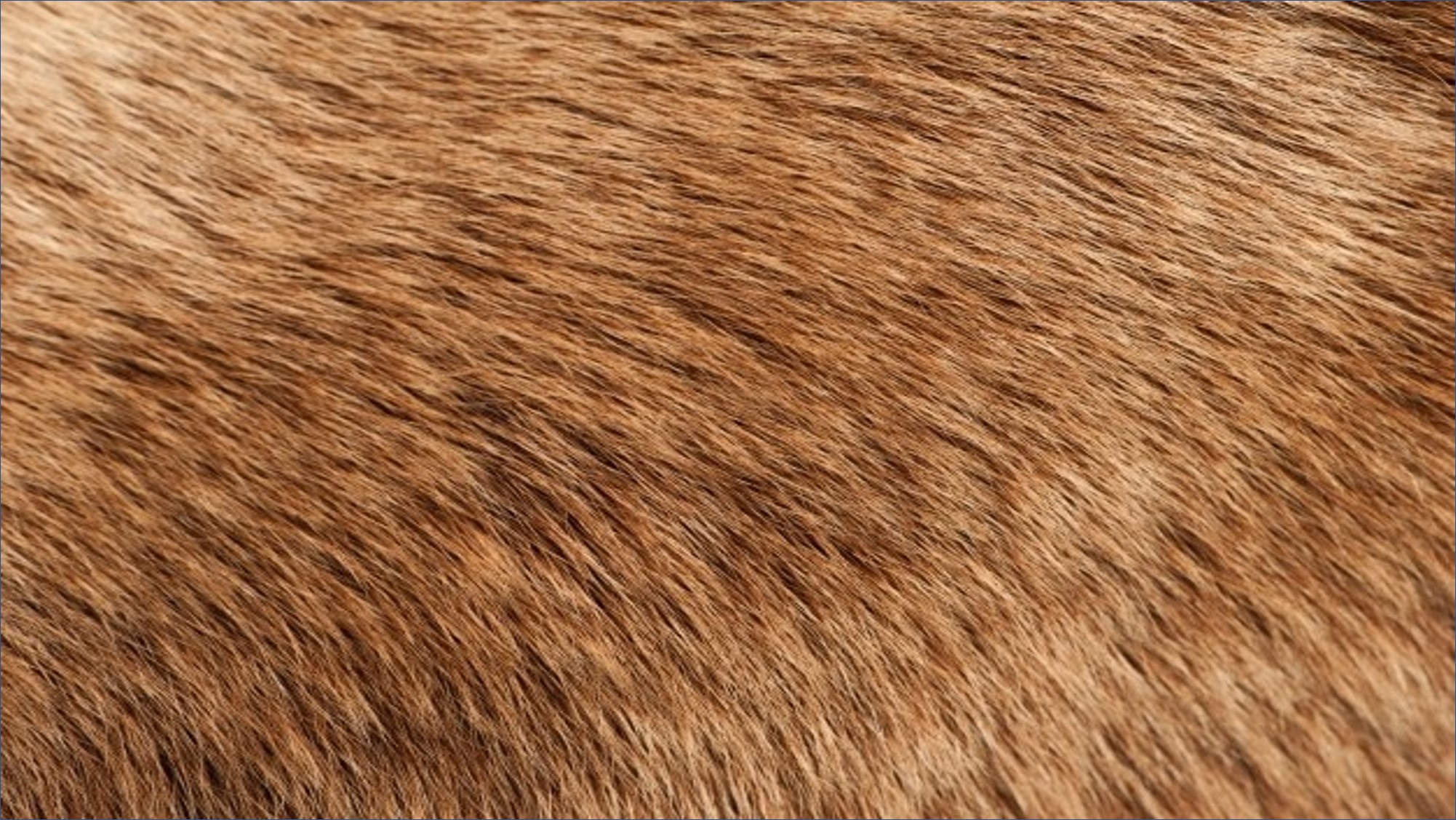}
  \end{minipage}
  \caption{
Experimental objects to give four different texture perceptions.
(a) fabric,
(b) glass,
(c) paper,
(d) fur.
}
\end{figure}

\begin{figure}[t]
\centerline{\includegraphics[width = \columnwidth]{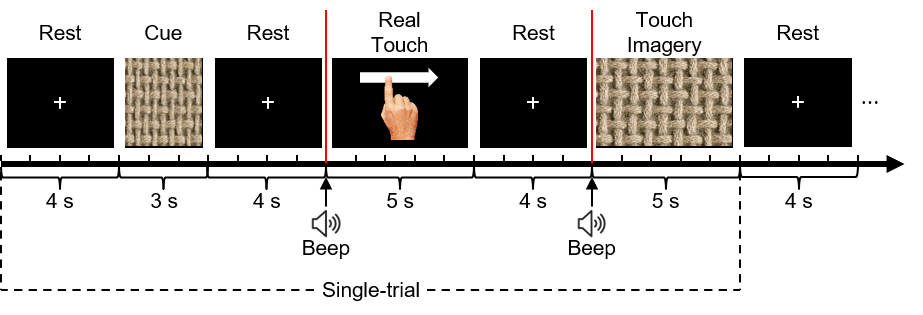}}
\caption{The paradigm of real touch and touch imagery experiment. The subject first performed a ``Real Touch'' session and then a ``Touch Imagery'' session. At the beginning of each session, an auditory cue of less than 0.5 seconds and a visual cue were given.}
\end{figure}

\subsection{Data Analysis}
Data analysis was performed offline using the BBCI toolbox and EEGLAB toolbox (version 14.1.2b). Filtered EEG signals were used to evaluate the performance of real touch stimuli and touch imagery. Raw EEG signal filtered from 1 to 45 Hz using a bandpass filter. The features of the EEG signal are extracted as a common spatial pattern (CSP). Alpha and beta band powers have been used in previous work on tactile image analysis. Therefore, the alpha and beta band energies were extracted and filtered by CSP. The linear discriminant analysis (LDA) was used as a classifier to classify real touch stimuli and touch imagery\cite{Greco2019IEEE}. The EEG signal was bandpass filtered at [1-50] Hz using a 3rd order Butterworth filter to analyze alpha and beta band power activity based on spatial information. The BBCI toolbox was used to visualize the topographic map of the preprocessed data. Real touch stimulus and spatial information of touch imagery data in 5 time zones: (0-1,000) ms, (1,000-2,000) ms, (2,000-3,000) ms, (3,000-4,000) ms, (4,000-5,000) ms Analyzed\cite{Kam2013Neuro, MLee2017SCIREP}. To understand the variance of power in the alpha and beta bands, we performed channel time-frequency of real touch stimulus and touch imagery data. Changes in spectral power with event-related changes at each time and at each frequency during the test were analyzed using the event-related spectral perturbation (ERSP) method. ERSP analysis was performed for frequencies ranging from 1 to 30 Hz for all channels using 200-time points. The baseline for calculating the ERSP was taken from the last 500 ms of the rest phase before the real touch stimulus and touch imagery phase\cite{kim2015neuro, kim2015neu}. The noise and artifacts effects induced by preliminary movements were eliminated before using the data for analysis. The data between 500-4,500 ms based on the onset point were used for the analysis.\\

\section {Results and discussion}

\subsection{Neurophysiology Analysis}

\begin{figure}[t]
\centerline{\includegraphics[width = \columnwidth]{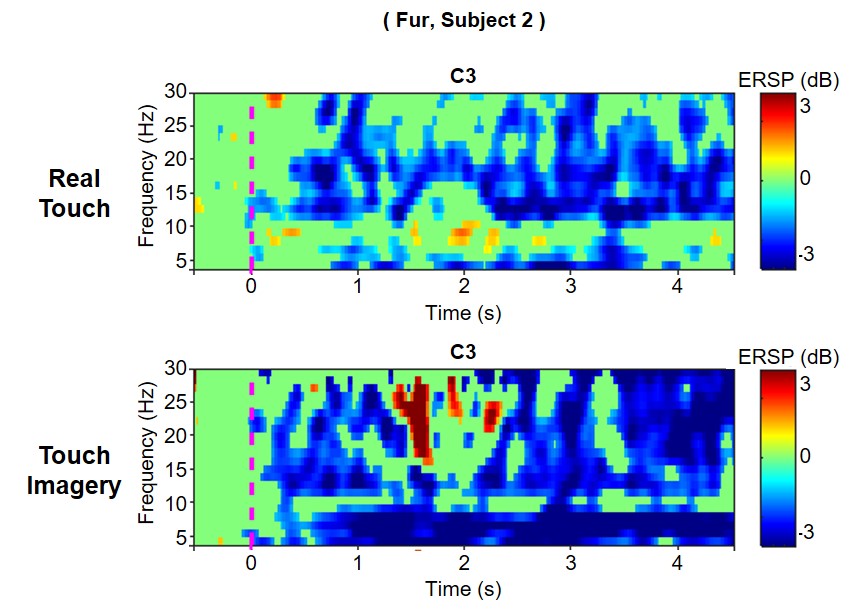}}
\caption{The event-related spatial perturbation of real touch and touch imagery. (Object: Fur, Subject 2 at C3 channel).}
\end{figure}
\begin{figure}[t]
\centerline{\includegraphics[width = \columnwidth]{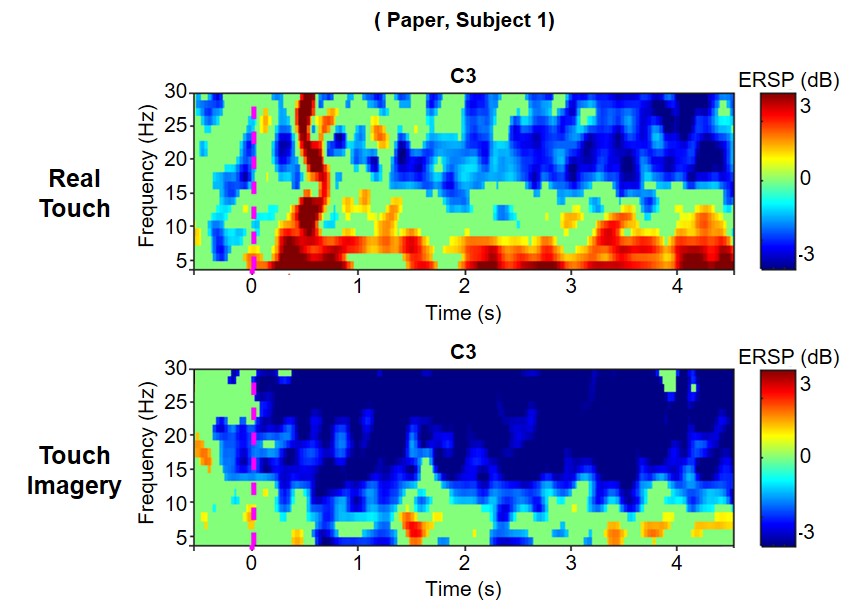}}
\caption{The event-related spatial perturbation of real touch and touch imagery. (Object: Paper, Subject 1 at C3 channel).}
\end{figure}
A comparison of the event-related spectral perturbation (ERSP) of real touch and touch imagery at the basic-FIR filtered C3 channel is shown in Fig. 4 and 5. For the real touch sessions, upon the onset of the tactile exploration of natural textures (on the right hand at 0 s), a continuous desynchronization is observed in the alpha-beta frequency band (8-30 Hz) for approximately 5 s. In addition, contralateral desynchronization is also observed in the low-alpha frequency band (8-10 Hz). Similarly, for the touch imagery sessions, a prominent and sustained desynchronization is observed in the alpha-beta frequency band (8-30 Hz) during imagining the tactile sensation.
The ERSPs in both sessions are very similar. We have observed a noticeable desynchronization in the alpha-beta frequency band (8-30 Hz). Various motor imagery studies have reported that desynchronization occurred in the alpha band (8-13 Hz). On the other hand, somatosensory studies reported that desynchronization was observed in the alpha-beta band. Based on the results of these many previous studies, it can be seen that our experimental results have considerable validity that can be explained by the EEG response in the somatosensory area to the sense of touch\cite{Ebrahim2017srep, Chen2019TNSRE, Schirmer2019Cortex}.

Another notable thing in our experimental results is that the brain-wave response of touch imagery is more prominent than that of real touch. This result is contrary to the general expectation that the actual contact response will be more pronounced than expected.

Most BCI studies on tactile sensation required a device for somatosensory stimuli, such as vibration or actual tactile contact. However, through our experimental results, we confirmed the possibility of using the sense of touch as a novel BCI paradigm through touch imagery without devices for stimulation.

\subsection{Performance Evaluation}

\begin{figure}[t]
  \centering
  \subfloat(a){
  \minipage{.2\textwidth}\centering\includegraphics[width=0.9\columnwidth]{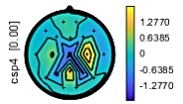}
  \endminipage}
  \subfloat(b){
  \minipage{.2\textwidth}\centering\includegraphics[width=0.9\columnwidth]{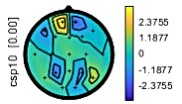}
  \endminipage}
  \\
  \subfloat(c){
  \minipage{.2\textwidth}\centering\includegraphics[width=0.9\columnwidth]{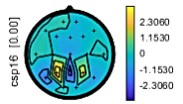}
  \endminipage}
  \subfloat(d){
  \minipage{.2\textwidth}\centering\includegraphics[width=0.9\columnwidth]{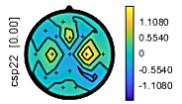}
  \endminipage}
  \caption{
Scalp-topoplot by CSP filter of subject 1 corresponding to four objects.
(a) fabric,
(b) glass,
(c) paper,
(d) fur.
}
\end{figure}
\begin{table}
\centering
\renewcommand{\arraystretch}{1.1}
\renewcommand{\tabcolsep}{5mm}
\begin{tabular}{ccc}
\multicolumn{3}{c}{T{A}{B}{L}{E} {I}}\\
\multicolumn{3}{c}{4-{C}lass {C}lassification {A}ccuracy (5-by-5-fold cross-validation)}\\
\multicolumn{3}{c}{}\\ \hline
& \textbf{Real touch} & \textbf{Touch imagery}\\ \hline
\textbf{Subject 1} & 70.95($\pm$1.50)\% & 56.25($\pm$1.53)\%\\
\textbf{Subject 2} & 65.20($\pm$2.53)\% & 58.85($\pm$1.43)\%\\
\textbf{Average} & 68.07($\pm$4.0)\% & 57.55($\pm$1.83)\%\\ \hline
\end{tabular}
\end{table}
In this preliminary experiment, brain signals related to real touch and touch imagery were measured for four different objects (fabric, glass, paper, and fur). We used the CSP for feature extraction from brain signals and the LDA for classification as described in section II.

Table I showed a comparison of the classification accuracies of real touch related brain signals and touch imagery related brain signals.
The performance of real touch was 68.07($\pm$4.0)\% and that of touch imagery was 57.55($\pm$1.83)\% respectively. And the results were over 25\% chance level in four classes. The performance according to the subjects was similar ($\pm$4.0\% and $\pm$1.83\%).
In touch imagery, the fabric classification accuracy was the highest at over 74.0\%, and the paper was the lowest at 52\%. Even if excluding the fact that this study was our first experiment on touch imagery and the classification by basic machine learning techniques, we got a fairly high classification performance compared to the chance level. Furthermore, we can expect higher performance when classifying measurement data using state-of-the-art techniques such as deep neural networks\cite{KPark2016Winter}.
We propose that touch imagery can provide new possibilities for BCI-based application systems.\\


\section{Conclusion and Future Works}
In this study, brain signals related to real touch and touch imagery were analyzed by brain region and frequency. And we classified the tactile-related brain signals for four different objects (fabric, glass, paper, and fur) based on real touch and touch imagery. The classification performance significantly exceeded the level of opportunity for all classes. And the possibility of touch imagery the possibility was demonstrated.
 Furthermore, this was our first experiment and we are going to implement a suitable deep learning architecture rather than machine learning for classification to further improve performance in the future.

In Fig. 4 and 5, we can see a similar aspect of related literature on somatosensory through the ERSP for real touch and touch imagery related EEG. We are going to analyze the neurophysiological analysis of what this means in brain signals related to touch imagery.

Through a questionnaire of the subjects, we found that the paradigm proposed in this study had a problem that the subject's concentration declined over time. Each session has 50 trials per class and subjects should perform a total of 400 trials in two sessions. In a BCI experiment, a more concise and effective experimental paradigm should be developed because a large number of trials and a long time can cause fatigue and decreased concentration in the subjects. We will improve the proposed paradigm to get better brain signals.\\

\section{Acknowledgement}
The authors thank to B.-H. Kwon, K.-H. Lee, B.-H. Lee, H.-J. Ahn and D.-H. Lee for help with the database construction and useful discussions of the experiment.\\
\bibliographystyle{IEEEbib}
\bibliography{refs}

\end{document}